\documentclass{article}
\usepackage{spconf,amsmath,graphicx}
\usepackage{siunitx}
\usepackage{hyperref}
\usepackage{cleveref} 

\usepackage{enumitem}
\setlist{nosep, leftmargin=14pt}

\usepackage{mwe} %

\newcommand{\FD}{\ensuremath{\text{FD}}}
\newcommand{\logFC}{\ensuremath{\log(\text{FC})}}

\title{Two-step registration method boosts sensitivity in \linebreak longitudinal fixel-based analyses}
\name{Aurélie Lebrun$^{1,2}$, Michel Bottlaender$^{1,2}$, Julien Lagarde$^{2,3,4}$, Marie Sarazin$^{2,3,4}$, Yann Leprince$^{1}$}

\address{\ninept$^{1}$UNIACT, NeuroSpin, CEA, Université Paris-Saclay, France;    
    $^{2}$BioMaps, SHFJ, CEA, CNRS, Inserm, Université Paris-Saclay, France;\\\ninept
    $^{3}$Department of Memory and Language, GHU Paris Psychiatry and Neurosciences, France;
    $^{4}$Université Paris-Cité, France}

\begin{document}
\maketitle
\begin{abstract}
Longitudinal analyses are increasingly used in clinical studies as they allow the study of subtle changes over time within the same subjects. In most of these studies, it is necessary to align all the images studied to a common reference by registering them to a template. In the study of white matter using the recently developed fixel-based analysis (FBA) method, this registration is important, in particular because the fiber bundle cross-section metric is a direct measure of this registration. In the vast majority of longitudinal FBA studies described in the literature, sessions acquired for a same subject are directly independently registered to the template. However, it has been shown in T1-based morphometry that a 2-step registration through an intra-subject average can be advantageous in longitudinal analyses. In this work, we propose an implementation of this 2-step registration method in a typical longitudinal FBA aimed at investigating the evolution of white matter changes in Alzheimer's disease (AD). We compared at the fixel level the mean absolute effect and standard deviation yielded by this registration method and by a direct registration, as well as the results obtained with each registration method for the study of AD in both fixelwise and tract-based analyses. We found that the 2-step method reduced the variability of the measurements and thus enhanced statistical power in both types of analyses.
\end{abstract}
\section{Introduction}
\label{sec:intro}
Longitudinal analyses are powerful experimental designs that limit inter-subject variability compared to cross-sectional analyses, allowing the study of more subtle effects. Such analyses often require all acquired images to be align to a common reference by registering them to a template.

In the study of brain white matter changes using diffusion magnetic resonance imaging (MRI), the recently developed fixel-based analysis (FBA) method \cite{Dhollander_2021} allows to overcome the major limitation of classical diffusion MRI models in the presence of the very frequent fiber crossing situations, by defining a new basic element, called fixel, which represents one fiber population within a voxel, and by performing analyses fixelwise instead of voxelwise. This method allows the study of white matter at both micro- and macroscopic scales through two metrics: the fiber density (FD), which describes the microstructure, and the fiber bundle cross-section (FC), which describes the macrostructure. This second metric is derived from the Jacobian of the deformation fields between the native images and a study-specific population template \cite{Dhollander_2021}. A typical FBA pipeline therefore requires the construction of such a template and a registration step of all acquired images to this template.

In the context of a longitudinal FBA, to the best of our knowledge, in the vast majority of the studies described in the literature, images acquired for a same subject are directly independently registered to the population template, as it is done in the usual FBA pipeline, and the longitudinal design is only addressed in the construction of an unbiased population template and in statistical analyses \cite{Genc_2018}. However, in longitudinal studies conducted with other imaging modalities, such as T1-based morphometry, it has been shown that also adapting the registration to the template step by performing a 2-step registration method can offer advantages, such as reducing the variability of the effects that are measured \cite{Reuter_2012}. This 2-step registration method consists of, first co-registering the images obtained for a same subject and computing an intra-subject average, and then registering this average to the population template. 

In this study, we hypothesized that this 2-step registration method could also be advantageous in the context of longitudinal FBA. We therefore propose an implementation of this method in a typical longitudinal FBA aimed at investigating the evolution of white matter changes in Alzheimer's disease (AD) which includes AD patients and healthy controls. 
To quantitatively compare this method with the usual direct registration method, we implemented two FBA pipelines that were identical except for the registration to the population template step, and we compared the mean rates of change of the fixel-based metrics and the standard deviation of these rates of change yielded by each registration method at the fixel level. We then compared the results obtained with each method when testing if the fixel-based metrics decrease over time in AD, in both fixelwise and tract-based analyses.

\section{Methods}
\label{sec:methods}
We included 31~subjects from the SHATAU7/IMATAU cohort (EudraCT: 2015-000257-20). All participants provided informed written consent for their participation. 16~participants (mean age: 72.6 (6.0); F = 9) were diagnosed with typical amnestic AD and 15~participants (mean age: 68.3 (3.6); F = 10) were cognitively healthy controls (HC). Descriptions of inclusion criteria are available in a previous publication \cite{Lebrun_2024}.

MRI data was acquired at the CENIR, Paris Brain Institute, using a 3-tesla Siemens Magnetom Prisma scanner and a 64-channel coil. All participants underwent two imaging sessions with a 2 (SD = $0.27$) years interval. At each session, diffusion images were acquired using a single-shot EPI sequence (TE = 77 \unit{\ms}, TR = 7 \unit{\s}, voxel size = 1.3 \unit{\mm} isotropic). Each diffusion acquisition comprised three shells of diffusion with $b$-values of 200, 1700, 4200 \unit[per-mode=symbol]{\s\per\mm\squared}, with 60 directions per shell, and three volumes without diffusion weighting. Other acquisition parameters and the usual preprocessing steps of the diffusion data are described in \cite{Lebrun_2024}.

\subsection{Fixel-based analysis pipeline}
We implemented a FBA pipeline following the recommendations of the authors of the method \cite{Dhollander_2021} unless otherwise stated, and using the MRtrix3 commands \cite{Tournier_2019}. 

Fiber orientation distributions (FODs) were obtained for each session of each subject using multi-shell multi-tissue constrained spherical deconvolution \cite{Jeurissen_2014}, and the mean three-tissue response functions of healthy controls \cite{Lebrun_2024}.

An unbiased longitudinal population FOD template was created using data of 28 subjects (14 AD patients, and 14 HC) that showed the best quality after visual assessment. For each of them, we computed an intra-subject FOD average using non-linear registration (see step 1 of the 2-step registration method in \autoref{sec:two_registration_methods}). We then used the 28 intra-subject FOD averages to generate the population FOD template. 

For each subject, we registered the FODs obtained for each session to the population FOD template using the two different methods that we aimed to compare (\autoref{sec:two_registration_methods}). For each method, we segmented the registered FODs into fixels and we derived the fixelwise \FD{} and \logFC{} metrics \cite{Dhollander_2021}. 

As longitudinal studies focus on the evolution of the metrics over time, we calculated an annualized rate of change $R$ of \FD{} and \logFC{} between sessions 1 and 2 as follows:
\begin{equation}
    R(\text{metric}) = \frac{\text{metric}_{\text{session 2}} - \text{metric}_{\text{session 1}}}{\text{time interval (years)}}
\end{equation}

\subsection{Two registration methods}
\label{sec:two_registration_methods}
As previously stated, we used two different registration methods to register each subject FODs to the population template.

\begin{figure}
    \centering
    \includegraphics[width=8.5cm]{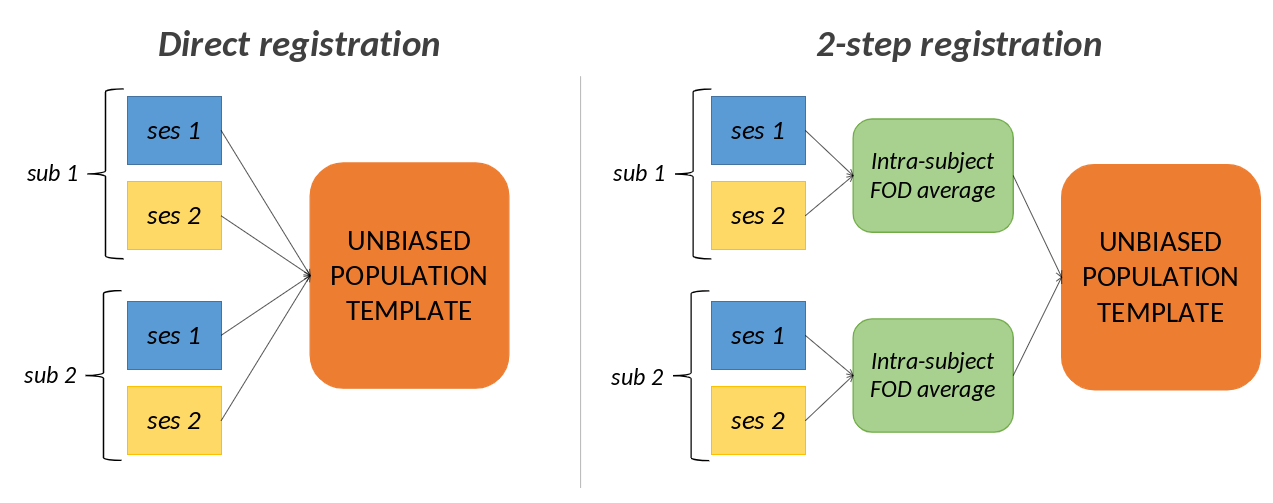}
    \caption{\ninept Illustration of the two registration methods: left, direct registration; right, 2-step registration.}
    \label{fig1}
\end{figure}

\noindent\textbf{Direct registration}\quad (\autoref{fig1}, left) FODs of each session of one subject are non linearly independently registered to the population template using the usual FOD-based registration \cite{Raffelt_2011} (MRtrix3 command 'mrregister', default parameters). %

\noindent\textbf{2-step registration}\quad (\autoref{fig1}, right) FODs of each session of one subject are, first, non-linearly co-registered, transformed to the midway space between the two sessions, and averaged to obtain an intra-subject FOD average (\textbf{step 1}). This intra-subject FOD average is then non-linearly registered to the population template (\textbf{step 2}). Both registration steps are performed using FOD-based registration \cite{Raffelt_2011} with default parameters. The two warps resulting from the two registration steps are then composed to generate a single warp per session,
which is reinjected in the usual FBA pipeline to transform the native FODs to the population template without performing multiple interpolations. 

\subsection{Comparison of the two registration methods}
We investigated the impact of each registration method on longitudinal FBA results obtained for $R(\FD)$ and $R(\logFC)$, first in fixelwise analyses, and then in tract-based analyses.

\noindent\textbf{Fixelwise analyses}\quad We generated a whole brain tractogram on the population template following the usual recommendations \cite{Dhollander_2021} to perform statistical tests using connectivity-based fixel enhancement (CFE) \cite{Raffelt_2015}.

For each metric (\FD{} and \logFC{}) and for each registration method, we computed for each fixel of the population template the mean rate of change $R$, and its standard deviation, within each group of participants (AD patients and HC). We used joint histograms of these fixelwise metrics and linear regressions, in order to quantitatively compare the results yielded by each registration method. For the standard deviation, we set the linear regression intercept to 0 to derive a percentage of change of the variability yielded by the 2-step method based on the estimation of the regression coefficient. 

Finally, for each registration method, we tested whether $R(\FD)$ and $R(\logFC)$ are negative among AD patients. Statistical inferences were performed using CFE with non-parametric permutations and assuming independent and symmetric errors. Significance of the results was assessed using family-wise error correction with a type I error rate of 5\%.

\noindent\textbf{Tractwise analyses}\quad We used TractSeg \cite{Wasserthal_2018} to segment 25 major white matter tracts on the population template. For each tract and each registration method, we calculated the mean $R(\FD)$ and $R(\logFC)$ for the whole tract by taking the average of each rate of change over all fixels associated with the tract, weighted by track density. Among AD patients, we then tested whether the mean $R(\FD)$ and $R(\logFC)$ in each tract are negative using a one-sample t-test.

\section{Results}
\label{sec:results}

\begin{figure}
    \centering
    \includegraphics[width=8.5cm]{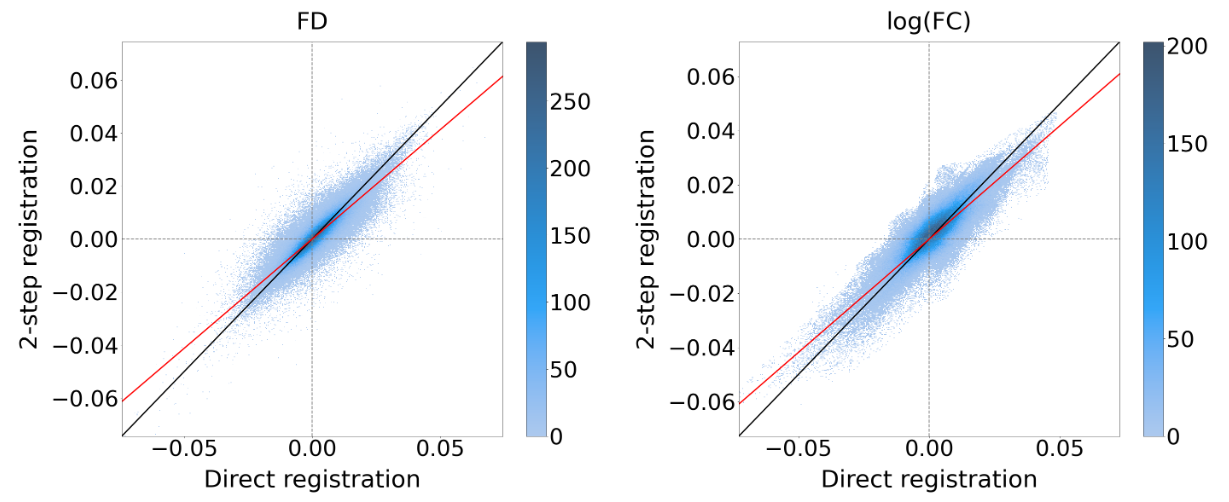}
    \caption{\ninept Joint histograms between the mean rate of change $R$ obtained for each fixel of the template among AD patients with the direct registration method vs.\ with the 2-step registration method. Black line: $y = x$, red line: linear regression.}
    \label{fig2}
\end{figure}

\noindent\textbf{Fixelwise analyses}\quad 
\autoref{fig2} shows that, among AD patients, the magnitude of the rates of change measured following the 2-step registration method is smaller on average compared to the one measured following the direct registration method, for both metrics \FD{} (regression coefficient = 0.82) and \logFC{} (regression coefficient = 0.84). Similar results were obtained for HC. Additionally, the intercepts of the linear regressions were close to 0 ($\leq 10^{-3}$), indicating that one method did not systematically over- or underestimate the mean $R$ compared to the other.

\begin{figure}
    \centering
    \includegraphics[width=8.5cm]{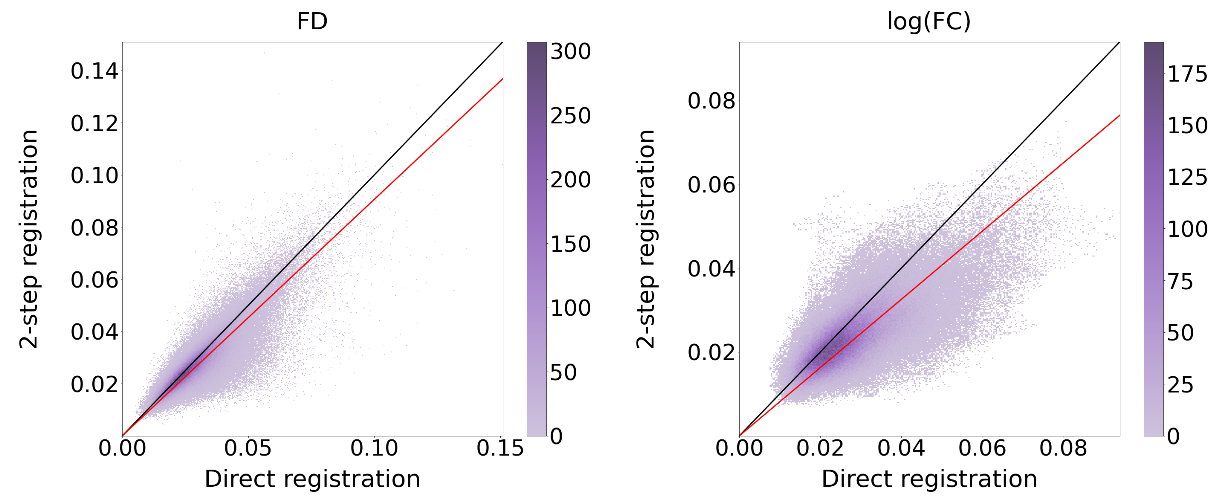}
    \caption{\ninept Joint histograms between the standard deviation of the rate of change $R$ obtained for each fixel of the template among AD patients, with the direct registration method vs.\ with the 2-step registration method. Black line: $y = x$, red line: linear regression with intercept set to 0.} 
    \label{fig3}
\end{figure}

\begin{figure}
    \centering
    \includegraphics[width=8.5cm]{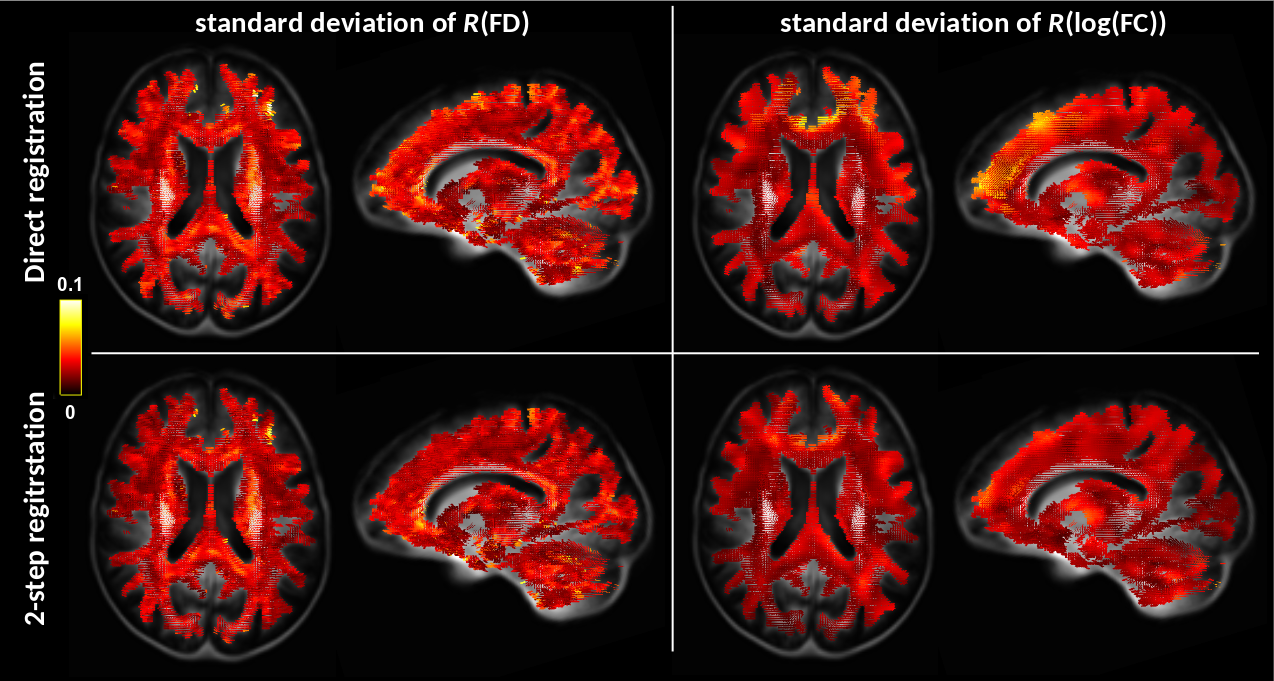}
    \caption{\ninept Standard deviation obtained for each fixel of the population template for $R(\FD{})$ (left) and $R(\logFC{})$ (right) among AD patients with the direct registration method (top) and the 2-step registration method (bottom).}
    \label{fig4}
\end{figure}

\autoref{fig3} shows that, among AD patients, the standard deviations of the rates of change measured following the 2-step registration method are smaller than those measured following the direct registration method for both metrics \FD{} and \logFC{}, this diminution being stronger for \logFC{}. Similar results were obtained for HC. We found that the 2-step method decreased the standard deviation of $R(\FD)$ by 9.4\% for AD patients, and by 10.2\% for HC, and that it decreased the standard deviation of $R(\logFC)$ by 18.6\% for AD patients, and by 13.9\% for HC. \autoref{fig4} shows that the white matter regions that exhibited the greatest reduction in variability with the 2-step method are located within the gyri, which are known to be more difficult to align.

\begin{figure}
    \centering
    \includegraphics[width=6.5cm]{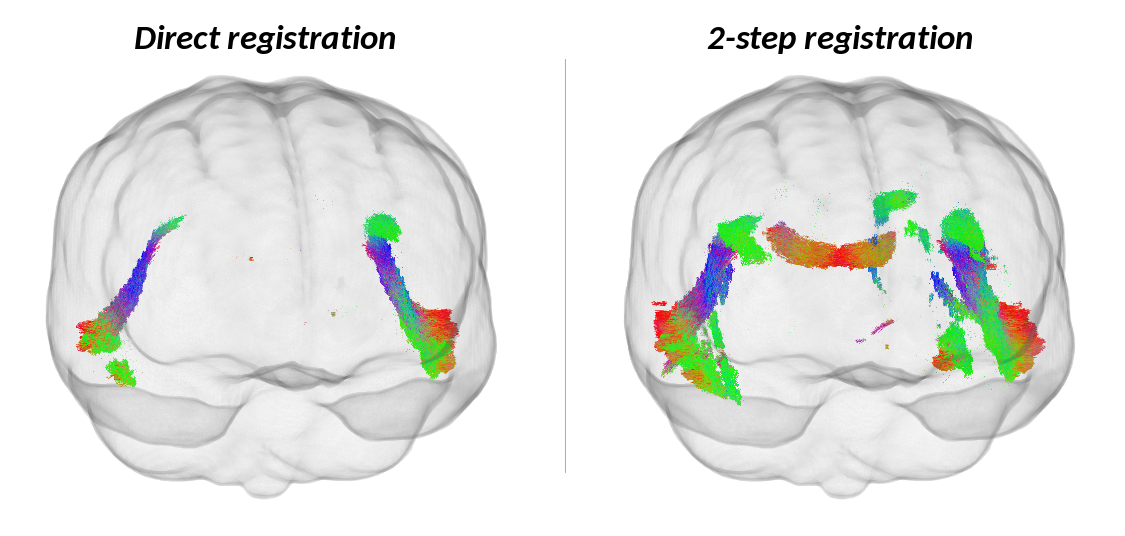}
    \caption{\ninept Significant results ($p < 0.05$ family-wise error corrected) obtained for the test `$R(\logFC) < 0$' among AD patients following each registration method.}
    \label{fig5}
\end{figure}

Finally, \autoref{fig5} shows that both registration methods produced significant results for the test `$R(\logFC{}) < 0$' that were located in the same brain regions, but the 2-step method allowed
to highlight more spatially extended results. We did not obtain significant results for the \FD{} metric.

\begin{figure}
    \centering
    \includegraphics[width=8cm]{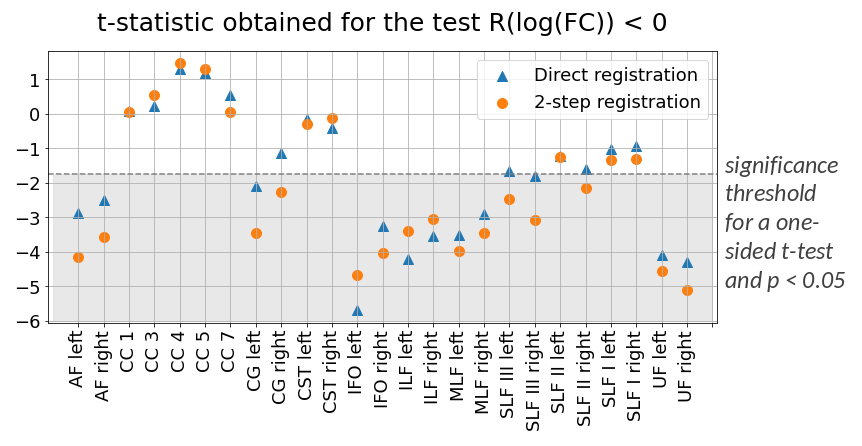}
    \caption{\ninept t-statistics obtained for the test `mean $R(\logFC) < 0$' among AD patients following each registration method in 25 tracts. See \cite{Wasserthal_2018} for the description of the tract acronyms.}
    \label{fig6}
\end{figure}

\noindent\textbf{Tractwise analyses}\quad \autoref{fig6} shows that the effects observed in fixelwise analyses in \autoref{fig5} persist in tract-based analyses, as the 2-step registration allowed to obtain more extreme t values, and therefore, more spatially extended results. For example, using this method, both left and right cingula (CG) and both left and right superior longitudinal fasciculi III (SLF III) are found significantly atrophied, instead of only the left CG and the right SLF III with direct registration.

\section{Discussion}
\label{sec:discussion}
In this study, we aimed to evaluate the impact of using a 2-step registration method in a longitudinal FBA pipeline rather than a direct registration. We found that this method reduced on average the mean absolute effects and, more importantly, the variability of the effects measured at the fixel level (between 9 and 18\%), ultimately increasing the effect sizes and allowing for more spatially extended significant results in fixelwise analyses. These effects persist in tract-based analyses. 

With direct registration, each session of a same subject is independently registered to the population template. This can lead to within-subject registration inconsistencies, as the registration of the different sessions of a same subject to the population template can converge to different local minima, especially in brain regions that are more difficult to align. Such inconsistencies can then affect the subsequently obtained metrics. The use of a 2-step registration method limits these risks, as the sessions of a same subject are first co-registered together before being registered to the population template using the same deformation field. 

This explains why, on average, the 2-step method reduced the mean absolute effect, without systematically underestimating it. Both registration methods yielded comparable mean absolute effects for the majority of fixels, but in brain regions that are more difficult to align and therefore more sensitive to registration inconsistencies, the direct registration method is likely to yield larger, more variable values. %

It is noteworthy that the sensitivity of both the FD and the FC analyses were improved by the 2-step registration method. If the effect of registration on the FC analysis is direct, since FC is derived from the registration, such benefits for the FD analysis are indirect evidence of a gain in registration accuracy. This influence of registration accuracy on FD has been discussed previously \cite{Dhollander_2021}. Regarding the FC metric, the 2-step method benefited the study of the AD patient group more than that of the control group. This can be explained by the fact that the patient group is more heterogeneous and more likely to contain images that are difficult to align.

This study demonstrates the advantage of using a 2-step registration method in a longitudinal FBA with two sessions per subject. In this context, the first step of the method was a co-registration of the two sessions to obtain a midway space between them. %
This 2-step method can be generalized to experimental designs where more than 2 sessions are acquired. In such situations, the midway space can be obtained by computing an intra-subject template with all acquired sessions \cite{Reuter_2012}. %

To conclude, this work demonstrates the feasibility and benefit of using a 2-step registration method in a longitudinal FBA, as this method reduces the variability of the effects being measured, thus enhancing statistical power. Conducting a similar study using a ground truth and simulated data could allow further investigation of the impact of this method. An application in the context of a longitudinal study of AD using FBA, with a comparison to healthy controls, is in preparation.

\ninept
\bibliographystyle{IEEEbib}
\bibliography{refs}

\section{Acknowledgments}
\label{sec:acknowledgments}
The authors have no conflicts of interest to report. We thank the clinical staff of the Hôpital Sainte-Anne and the CENIR for taking care of the subjects and performing the acquisition. We thank the French Health Ministry (PHRC-2013-0919 and PHRC- 0054-N 2010), CEA, \emph{Fondation Recherche Alzheimer}, \emph{Institut de Recherches Internationales Servier}, \emph{France-Alzheimer}, and \emph{Institut Roche de Recherche et Médecine Translationnelle} for their financial support.

\section{Compliance with ethical standards}
\ninept
All procedures performed in studies involving human participants were in accordance with the ethical standards of the institutional and/or national research committee.

\end{document}